\UseRawInputEncoding 
\documentclass[review,authoryear]{elsarticle}
\usepackage{lineno,hyperref}\modulolinenumbers[5]

\usepackage{amsmath}
\usepackage{multicol}
\usepackage{subfigure}
\usepackage{blindtext}
\usepackage{tabularx}
\usepackage[linesnumbered,ruled,vlined]{algorithm2e}
\usepackage{algpseudocode}
\usepackage{enumitem}
\usepackage{rotating}
\usepackage{pdflscape}
\usepackage{adjustbox}
\usepackage{amsmath}
\usepackage{amssymb}
\usepackage{mathtools}
\usepackage{pdfpages}

\DeclareMathOperator*{\argmax}{arg\,max}

\journal{Machine Learning with Applications}

\biboptions{round,sort,authoryear}

\begin{document}

\begin{frontmatter}

\title{Exploratory Data Analysis for Airline Disruption Management \tnoteref{mytitlenote}}
\tnotetext[mytitlenote]{This article represents one full chapter from the corresponding author's completed doctoral dissertation.}

\author[1]{Kolawole Ogunsina\corref{cor1}%
\fnref{fn1}}
\ead{kolawole08@gmail.com}

\author[2]{Ilias Bilionis\fnref{fn2}}
\ead{ibilion@purdue.edu}

\author[3]{Daniel DeLaurentis\fnref{fn3}}
\ead{ddelaure@purdue.edu}

\cortext[cor1]{Corresponding Author}
\fntext[fn1]{School of Aeronautics and Astronautics, Purdue University, United States.}
\fntext[fn2]{Department of Mechanical Engineering, Purdue University, United States.}
\fntext[fn3]{School of Aeronautics and Astronautics, Purdue University, United States.}

%\newpageafter{author}

%\author[Purdue University]{Elsevier Inc}

%\author[mysecondaryaddress]{Purdue University Thesis Office\corref{mycorrespondingauthor}}
%\cortext[mycorrespondingauthor]{Kolawole Ogunsina}
%\ead{kolawole08@gmail.com}


\begin{abstract}
Reliable platforms for data collation during airline schedule operations have significantly increased the quality and quantity of available information for effectively managing airline schedule disruptions. To that effect, this paper applies macroscopic and microscopic techniques by way of basic statistics and machine learning, respectively, to analyze historical scheduling and operations data from a major airline in the United States. Macroscopic results reveal that majority of irregular operations in airline schedule that occurred over a one-year period stemmed from disruptions due to flight delays, while microscopic results validate different modeling assumptions about key drivers for airline disruption management like turnaround as a Gaussian process.  
\end{abstract}


\end{frontmatter}

%\linenumbers

\section{Introduction}

Despite overcoming numerous financial and technical challenges over the last century through continued drive towards innovation and productivity, a complete solution to irregular operations in the airline industry has remained elusive. A major driver that has significantly stunted the progress in developing a full solution for airline disruption management is poor data integration and integrity.

Internal data containing real-time information about an airline's resources and their scheduled utilization over time, and external data for factors such as current and future weather forecasts, competitor activities, and air traffic control, are necessary for efficient operations \citep{SabreAirlineSolutions2013}. These bits of information and data must be readily available and accessible to represent drivers and constraints for scenarios induced by irregular operations, so as to facilitate the development of effective self-governing platforms for airline disruption management. In addition, whenever a new airline system is replaced or upgraded, new data sources are typically integrated into the existing framework \citep{Gershkoff2016}. The new data must be maintained for both existing and new applications, and thus present cost-intensive challenges for mitigating disruption because many facets of the airline infrastructure are impacted.   
\subsection{The Problem}
While there have been consistent improvements to the existing decision support systems used by human controllers in the Airline Operations Control Center (AOCC), two factors have continued to limit the performance of the disruption resolutions that are applied. First, the decision support systems do not explicitly proffer solutions to specific schedule disruptions. As such, human controllers in the AOCC are required to be reactive in addressing disruptions by using their best judgement based upon their prior experience from resolving the same (or similar) disruption. Second, majority of the decision support (computer) systems used by multiple departments (including the AOCC) in an airline and other air transportation stakeholders (e.g. airports) are not designed or developed at the same time nor by the same vendor \citep{SabreAirlineSolutions2013}. As such, information and data are required to be entered into multiple computer systems thereby exposing human controllers in the AOCC to data input problems and errors. Therefore, information entered into a decision support system for disruption management may be out of sync with other systems and yield incorrect decisions due to lack of data integrity. 

Furthermore, extant literature on the adoption of machine learning for decision support systems have only focused on addressing irregular operations for system-wide disruption management in the national airspace system as a whole. In other words, existing literature address irregular operations for system-wide disruption management by exploiting airport data from multiple airlines in the air transportation system. For instance, \cite{Liu2019} used machine learning to analyze air traffic management actions for incidences of ground delay programs across major airports in the United States. The authors employed a two-stage framework, such that the first stage used support vector machines to correlate incidences of ground delay programs with regional convective (weather) activity while the second stage trained logistic regression and random forest models by using a weather score obtained from the first stage. In addition, \cite{Ye2020} presented an approach for estimating aggregate flight departure delay at airports in China through supervised learning models. They applied four separate types of airport-related aggregate characteristics (including weather) to predict expected departure delays at a major airport in China, by employing linear regression, support vector machines, randomized trees, and LightGBM. While the models presented in both research studies provided acceptable predictive capacity for disruption management at the airport level in the respective air transportation networks, they (i.e. the models) do not describe the disruption management proclivities of a decision support system commissioned by a specific airline in the air transportation network. 

In the bid to improve data integrity and fidelity for existing decision support systems, airlines have significantly invested in creating better localized data collection platforms within their respective organizations, which can amass information from different sources within and outside the organization that is easily accessible through a centralized data server \citep{Amadeus2016}. As such, there is a need to fully leverage the ubiquity and accessibility of information (data) collected by existing platforms in the AOCC to enhance agile decision-making capabilities of the AOCC during airline disruption management. To that effect, this paper provides a comprehensive discussion on exploratory analysis administered on historical scheduling and operations recovery data supplied by a major airline in the United States. This exploratory analysis serves as the basis for the development of credible predictive and prescriptive models for airline disruption management.

\subsection{Contributions}
To the best of our knowledge, we provide the first literature that strictly employs the statistics of big data and machine learning to demonstrate the characterization and evaluation of a functional role (i.e. domain) in the AOCC of a major U.S. airline for disruption management. Thus, in contrast and complement to existing literature, our work provides research on irregular operations and disruption management accustomed to a specific U.S. airline based upon data provided by the airline. To achieve the aforementioned objectives, this paper enhances prior research and literature on irregular operations and airline disruption management through the following contributions:  
\begin{enumerate}
    \item We introduce and explore several abstraction methods for applying, enhancing, and sequestering raw features and labels in a historical data set on airline scheduling and operations recovery from a major U.S. airline, to readily identify relevant cognitive patterns for key drivers during airline disruption management. 
    \item We investigate the application of appropriate machine learning techniques for revealing patterns, pertinence, and properties of abstracted data features, which provide necessary a priori information (i.e. beliefs) for Bayesian and pseudo-Bayesian methods. These methods are subsequently employed in future work for developing functional models in an intelligent multi-agent system for airline disruption management. 
\end{enumerate}

\subsection{Paper Organization}
The next section in this paper provides an overview of the elements of historical scheduling and operations recovery data retrieved from a major U.S. airline, followed by a section that expansively discuses several relevant and interrelated processes for exploratory data analysis. We conclude with a summary of pertinent findings and areas for further research in Section \ref{conclusion}.

\section{Data Overview}\label{data_overview}
The raw data utilized for demonstrating the exploratory analysis discussed in this paper was provided by Southwest Airlines. Like many major U.S. airlines, Southwest Airlines employs an integrated AOCC organization wherein all functional roles share the same physical space (at the airline's headquarters in Dallas) and are hierarchically dependent on AOCC supervisors for multiple problem dimensions in airline operations recovery \citep{Deloitte2017}. As the largest carrier in the United States in terms of originating domestic passengers boarded with more than 4,100 flight schedule operations daily to over 100 destinations, the supervisors (and controllers) at Southwest Airlines Network Operations Control (SWA-NOC) seek to use technology to see the impact of their decisions to make better ones for improved disruption management. For many years, the controllers at SWA-NOC relied on gut instincts to track and understand how their disruption resolution actions cascaded throughout the airline's network, but could not inform their instincts with data. To address this issue, upper management at SWA-NOC created the Baker workgroup; an integrated team of supervisors and software developers dedicated to improving decision-making during disruption management by developing and enhancing a suite of computerized decision support systems called the Baker tool. In order to better support the Baker tool, the workgroup created an autonomous data collection platform to record flight schedules that are subject and not subject to different disruption incidents in the Southwest Airlines route network.

\begin{sidewaystable}
\begin{center}
\caption{Disruption outlook for functional domains in Southwest Airlines network operations control from September 2016 to September 2017} \label{tab:func_domain}
\begin{tabularx}{\linewidth}{ 
  | >{\centering\arraybackslash} p{0.2\textwidth}
  | >{\centering\arraybackslash} p{0.2\textwidth}
  | >{\centering\arraybackslash} p{0.2\textwidth}
  | >{\centering\arraybackslash} X
  | >{\centering\arraybackslash} X
  | >{\centering\arraybackslash} X | }
  \hline
	\textbf{Functional Domain} & \textbf{Affected Problem Dimension} & \textbf{Disruption Class} &\textbf{Delayed Flight Schedule Instances} & \textbf{Cancelled Flight Schedule Instances} & \textbf{Diverted Flight Schedule Instances}\\ \hline
	\textbf{\textit{Customer Hold}} & Aircraft and Passenger & Controllable & 46,870 & 0 & 0\\ \hline
	\textbf{\textit{Dispatch CSC}} & Aircraft and Crew & Controllable & 17,468 & 0 & 0 \\
	\hline
	\textbf{\textit{Flight Operations}} & Crew & Controllable & 36,370 & 1,099 & 909\\
	\hline
	\textbf{\textit{Fuel Management}} & Aircraft & Controllable & 4,841 & 0 & 0\\
	\hline
	\textbf{\textit{Ground Operations}} & Aircraft and Passenger & Controllable & 168,375  & 2518 & 460 \\ \hline
	\textbf{\textit{Inflight}} & Crew & Controllable & 79,444 & 984 & 67\\
	\hline
	\textbf{\textit{Maintenance}} & Aircraft & Controllable & 33,518 & 55 & 0 \\
	\hline
	\textbf{\textit{NAS}} & All & Uncontrollable & 22,644 & 2,619 & 433\\
	\hline
	\textbf{\textit{Security}} & Passenger & Controllable & 2,955 & 6 & 11\\
	\hline
	\textbf{\textit{Technology}} & All & Controllable  & 8,953 & 0 & 0 \\
	\hline
    \textbf{\textit{Weather}} & All & Uncontrollable & 12,659 & 12,156 & 4905\\
    \hline
\end{tabularx}
\end{center}
\end{sidewaystable}

Thus, the raw data generously provided to us for the research discussed in this paper contains approximately 1.1 million instances of direct flight schedules from Southwest Airlines route network operations recorded from September 2016 to September 2017; of which there are 620,000 flight schedules that were not subject to disruptions, over 430,000 flight schedules that were subject to flight delays, and approximately 26,000 flight schedules that were either cancelled or diverted. The instances of disrupted flight schedules (i.e. delayed, diverted and cancelled flight schedules) are distributed across eleven separate functional roles in SWA-NOC (i.e. the AOCC) that represent primary disruption resolution domains for aircraft, crew, and passenger problem dimensions in airline disruption management. Table~\ref{tab:func_domain} reveals a list of functional disruption resolution domains in the Southwest Airlines Network Operations Control, including the corresponding problem dimensions they seek to address, and the class of disruption and the number of instances of different effects of a disruption class for a specific functional domain. The disruption class defines the origination of a specific disruption, and as such, disruptions resolved by functional domains in SWA-NOC with a ``controllable" disruption class indicate that all instances of disrupted flight schedules associated with those domains were caused (or could have been avoided) by the airline. Conversely, disruptions resolved by functional domains in SWA-NOC with an ``uncontrollable" disruption class indicate that all instances of disrupted flight schedules affiliated with those domains were not caused (nor could have been avoided) by the airline. A brief description of the functional disruption resolution domains (or roles) in SWA-NOC highlighted in Table~\ref{tab:func_domain} is as follows:
\begin{enumerate}
    \item \textbf{\textit{Customer Hold}}: This functional domain addresses disruptions related to holding aircraft for passengers on inbound flight connections and holding aircraft to accommodate passengers off cancelled and delayed flights. As such, the customer hold functional domain resolves the aircraft and passenger problem dimensions in airline disruption management. Disruption instances for the customer hold domain accounted for about 11\% of delayed flight schedules in the Southwest Airlines route network over the one-year period (i.e. September 2016 to September 2017).
    \item \textbf{\textit{Dispatch CSC}}: This functional domain manages disruptions related to flight dispatch activities by the airline that also includes holding flights to accommodate international flight schedule slot times. To that effect, the Dispatch CSC functional domain addresses the aircraft and crew problem dimensions during disruption management, and disruption instances related to Dispatch CSC represented 4\% of delayed flight schedules in the airline operations between September 2016 and September 2017. 
    \item \textbf{\textit{Flight Operations}}: This functional domain resolves disruptions defined by Pilot (cockpit crew) scheduling activities as they relate to Pilot tardiness and normal aircraft readiness, and addresses the crew problem dimension of airline disruption management. Between September 2016 and September 2017, disruption instances related to Flight Operations represented about 8.5\% of delayed flight schedules, 5.7\% of cancelled flight schedules, and 13.5\% of diverted flight schedules in Southwest Airlines operations.
    \item \textbf{\textit{Fuel Management}}: This functional role in SWA-NOC manages disruptions related to aircraft fueling and other energy administration activities, and addresses the aircraft problem dimension during disruption management. Disruption instances related to Fuel Management between September 2016 and September 2017 represented 1.1\% of delayed flight schedules in Southwest Airlines operations.
    \item \textbf{\textit{Ground Operations}}: This functional domain in SWA-NOC manages disruptions defined by several activities ranging from passenger boarding and aircraft provisioning to ramp services and aircraft towing, and as such, resolves the aircraft and passenger problem dimensions in airline disruption management. Over the one year period of airline operations, disruptions related to Ground Operations accounted for the largest percentage of total flight schedule delays of 39\%, and the third highest percentage of total flight schedule cancellations of 13\%.
    \item \textbf{\textit{Inflight}}: Similar to Flight Operations, Inflight resolves disruptions defined by Flight Attendant (cabin crew) scheduling activities as they relate to Flight attendant tardiness and normal aircraft preparedness, and thus addresses the crew problem dimension of airline disruption management. Between September 2016 and September 2017, disruption instances related to Flight Operations represented about 18.5\% of delayed flight schedules, 5.1\% of cancelled flight schedules, and about 1\% of diverted flight schedules in Southwest Airlines operations.
    \item \textbf{\textit{Maintenance}}: This functional domain resolves disruptions defined by aircraft maintenance and inspection activities, and as such, addresses the aircraft problem dimension of airline disruption management. Disruption instances related to Maintenance represented 7.8\% of delayed flight schedules and about 0.3\% of cancelled flight schedules during Southwest Airlines operations from September 2016 to September 2017. 
    \item \textbf{\textit{NAS}}: This adopted functional role in SWA-NOC manages disruptions defined by air traffic control activities related to gate hold for congestion at departure and arrival airport stations. As such, the NAS functional domain addresses uncontrollable disruptions representing all problem dimensions during airline disruption management. Disruption instances associated with NAS represented 5.3\% of delayed flight schedules, 13.5\% of cancelled flight schedules and 6.4\% of diverted flight schedules during Southwest Airlines operations from September 2016 to September 2017.
    \item \textbf{\textit{Security}}: This functional domain addresses disruptions defined by security measures enforced to ensure the safety and convenience of passengers at airports prior to aircraft boarding. Its responsibilities includes managing disruptions due to baggage screening by TSA (Transportation Security Administration) at the skycap or ticket counter. As such, the Security functional domain resolves the passenger problem dimension during airline disruption management. Between September 2016 and September 2017, disruption instances related to Security represented the least percentages of total delayed, cancelled, and diverted flight schedules of 0.7\%, 0.03\%, and 0.16\%, respectively, in Southwest Airlines operations. 
    \item \textbf{\textit{Technology}}: This functional role manages all disruption activities defined by system-wide technology outages, and thus aims to resolve all problem dimensions during airline disruption management. Disruption instances related to Technology accounted for 2.1\% of all delayed flight schedules in Southwest Airlines operations between September 2016 and September 2017.
    \item \textbf{\textit{Weather}}: Similar to NAS, this adopted functional domain in SWA-NOC manages all kinds of uncontrollable disruption defined by inclement weather activities. To that effect, the Weather functional role aims to resolve the aircraft, crew and passenger problem dimensions during disruption management. Disruption instances associated with the Weather functional domain accounted for the highest percentage of cancelled and diverted flight schedules (62.6\% and 72.8\% respectively) among all functional domains in SWA-NOC between September 2016 and September 2017. In addition, delayed flight instances related to Weather represented 2.9\% of the total delayed flight instances addressed by all functional domains in SWA-NOC over the one year data collation period. 
\end{enumerate}

\begin{figure}[htb!]
\centering
\includegraphics[width=0.9\textwidth]{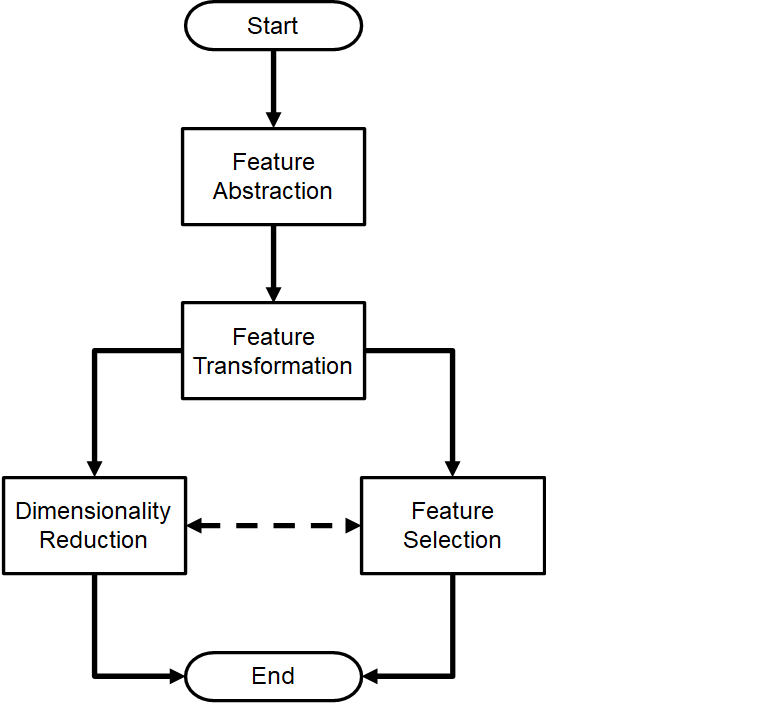}
\caption{Procedure for exploratory data analysis in a data-driven paradigm for airline disruption management}
\label{fig:Data_Analysis}
\end{figure}

\section{Data Analysis}
The previous section provided a macroscopic overview of disruption activities for different functional roles in SWA-NOC and revealed that about 42\% of all flight schedules for Southwest Airlines route network operations from September 2016 to September 2017 were disrupted. As a result, the functional disruption resolution domains in SWA-NOC were most likely to address irregular operations due to delayed flight schedules, which represent approximately 94\% of all disrupted flight schedules recorded from September 2016 to September 2017. Furthermore, there are two separate chunks of data which are defined by the occurrence of disruption during flight schedule execution. The first chunk, which is known as the \textit{non-disrupted} data set, represents the larger chunk that contains instances of flight schedules that executed without any disruption. The smaller chunk, also known as the \textit{disrupted} data set, contains instances of flight schedules that executed with disruption, and thus represent instances of flight schedule execution due to irregular operations. As such, the major difference between the \textit{non-disrupted} data set and \textit{disrupted} data set is the existence of additional data features (i.e. disruption features) in the \textit{disrupted} data set that indicate different types of disruption. However, there is a small subset (6\%) of the \textit{disrupted} data set that represents instances of canceled and diverted flight schedules, which have less data fields with sparse data entries that present significant challenges for machine learning applications. To that end, we restrict the scope of the research presented in this paper to irregular operations based upon delayed flight schedules and ignore flight cancellations and diversions which are primarily limited to the Weather functional domain. Henceforth, irregular operations only represent controllable and uncontrollable disruptions due to delayed flight schedules.

Fine details that highlight pertinent high-level patterns for elements (or features) that define each flight schedule in the raw data set introduced in Section~\ref{data_overview} can not be readily observed nor acknowledged through macroscopic inspection. Hence, in order to mine relevant microscopic information from raw data, this section elucidates the exploratory data analysis used to effectively generalize pattern-finding schemes for consistent flight schedule features that are applicable in all functional roles in SWA-NOC. Fig~\ref{fig:Data_Analysis} shows the general procedure adopted for performing exploratory data analysis on the raw data set. The process commences by abstracting separate data features that represent distinct properties of flight scheduling and operations, next raw data features are transformed into data forms that are readily decipherable by appropriate machine learning algorithms. The data transformation is necessary for applying separate methods for identifying critical data features and reducing the dimension space of the data set achieved by the feature selection and dimensionality reduction processes shown in Fig~\ref{fig:Data_Analysis}. The subsequent parts of this section provide more insight into the aforementioned processes as they relate to the historical scheduling and operations data adapted for the microscopic analysis techniques discussed in this paper.  

\subsection{Feature Abstraction}

\begin{figure}[t!]
\centering
\includegraphics[width=0.99\textwidth]{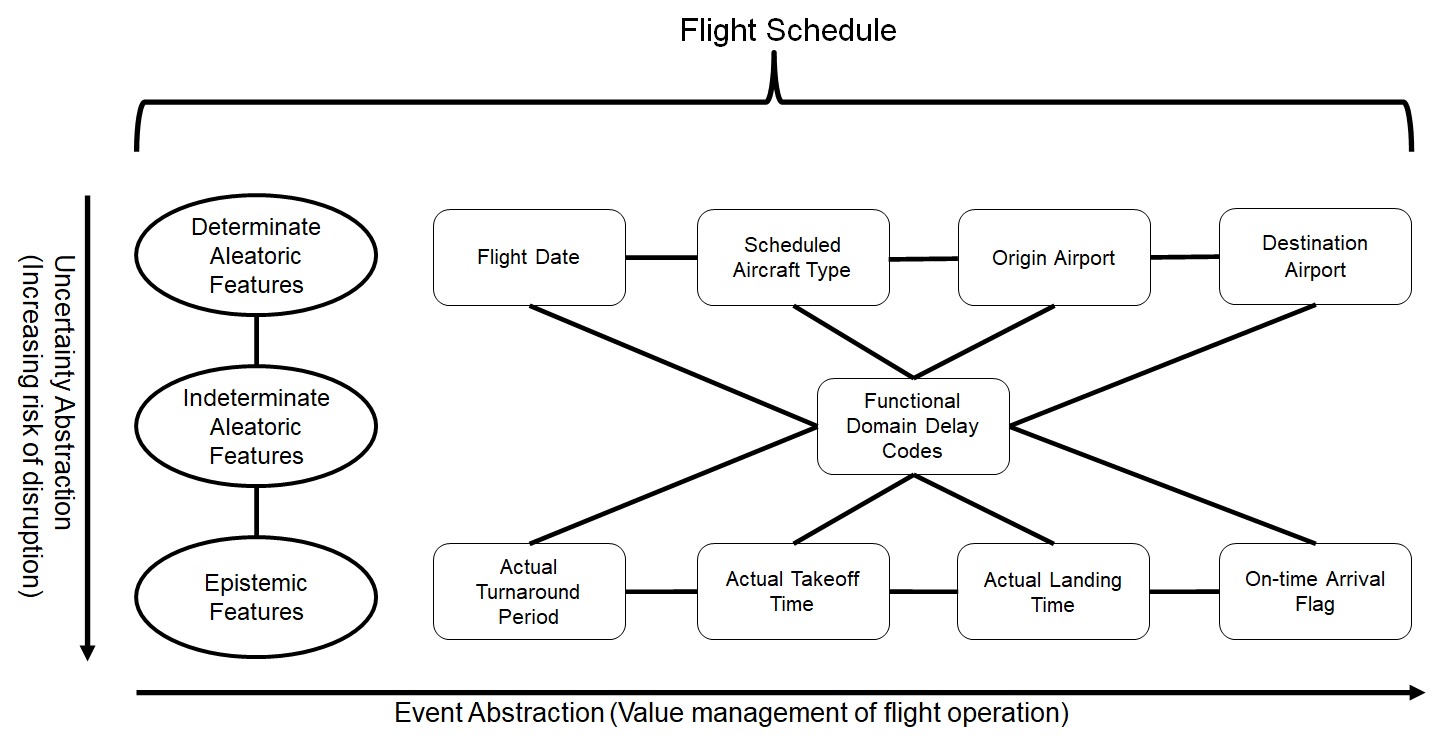}
\caption{Sample knowledge abstraction of a basic flight schedule for airline disruption management}\label{fig:KAH}
\end{figure}

Feature abstraction (often referred to as data abstraction) is an effective technique for accommodating semantic relationships between features in a database \citep{Ramakrishnan1995}. Feature abstraction ensures that features that define specific properties of fundamental tenets of flight scheduling and operations (embedded in each flight schedule from the raw data set) is generalized into abstract values \citep{Huh2000}. As such, a specific property can be viewed as a specialized quality of an abstract value. The properties describing the tenets of flight scheduling and operations are described by two separate principles of abstraction that represent a knowledge abstraction \citep{Abbott1987, Liskov1988, Huh2000}. These related feature abstraction principles exist for creating semantic associations among data features for airline scheduling and disruption management and are namely: \textit{event abstraction} and \textit{uncertainty abstraction}. 
\begin{itemize}
    \item \textit{Event abstraction}: Event abstraction serves two primary purposes. First, it characterizes the importance of planned airline activities and resources associated with a specific flight schedule, and how their importance can be subject to change prior to (or on) day of operation due to the risk of disruption from passenger-boarding at a departure station to aircraft gate-parking at an arrival station. Second, event abstraction defines the manner in which planned airline activities for a specific flight schedule varies during schedule execution based upon the impact of irregular operations. As such, the event abstraction principle is synonymous to \textit{flight operation value abstraction}, wherein the specific value (or profit) that a particular flight schedule in the airline route network provides is appraised by how effectively the flight schedule features estimated prior to schedule execution align with flight schedule features that are realized during and after schedule execution. To that effect, flight schedule features that are determined prior to schedule execution represent low-value features for event abstraction, and flight schedule features estimated during and after schedule execution represent high-value features for event abstraction.
    
    \item \textit{Uncertainty abstraction}: From an airline planning perspective, the uncertainty abstraction principle, also analogous to a \textit{functional domain planning abstraction}, defines the manner in which the uncertainty for the risk of irregular operations during schedule planning and execution is quantified and propagated via various features in the data set. Most airlines, including Southwest Airlines, adopt a perspective on the scheduling process that accentuates the internal planning approach of different planning departments as an iterative cycle of flight schedule development and assessment over a timeline horizon, such that the flight schedule is continually adjusted and optimized until a suitable schedule is obtained or the planning period is over \citep{Grosche2009}. Thus, the primary purpose of \textit{uncertainty abstraction} is to express the relationships among flight schedule features that are representative of the transitions through three iterative and interconnected airline planning phases namely: strategic planning, tactical planning and operational planning respectively \citep{Grosche2009b}. Strategic planning, otherwise known as ``future scheduling", focuses on long-term decision-making for the subsequent tactical and operational planning phases, such that a generic service plan consisting of essential and viable sets of serviceable routes without specific aircraft and crew assignments and tentative departure and arrival times are determined. Tactical planning or ``current scheduling" focuses on creating a refined schedule of operations for the service plan based upon the resources that are actually expected to be available to the airline over a definitive time period. Lastly, the operational planning phase focuses on adjusting the schedule generated in the tactical planning phase with respect to changes in demand for air travel prior to executing the flight schedule, and also on executing the schedule with minimal penalty (cost) in the event of unexpected disruptions on the day of operation (i.e. rescheduling for disruption management) \citep{Mathaisel1997}. 
\end{itemize}
By applying the event and uncertainty abstraction principles, we identify three separate classes of features in the raw data that can be used to enable robust airline disruption management and are described as follows:

\begin{enumerate}
    \item \textit{Determinate aleatoric features}: These represent flight schedule features that are determined during the strategic planning phase of airline planning and are required to remain unchanged during schedule execution on the day of operation. Examples of determinate aleatoric features include flight date, origin (departure) station, destination (arrival) station, route originator indicator, route distance, etc. With respect to disruption management, determinate aleatoric features represent flight schedule features whose alternatives do not differ considerably each time the AOCC invokes a disruption management initiative. Thus, from a statistical perspective, determinate aleatoric features are flight schedule features that are subject to the least possible uncertainty for the risk of reassessment (or alteration) during irregular operations for disruption management, based upon inherent randomness of disruption events. For instance, airport identifiers and exact longitude and latitude coordinates that provide specific information for origin and destination stations are always assumed to remain unchanged, by the AOCC, during the recovery of a delayed flight schedule. However, if a human specialist in the AOCC chooses to divert the same delayed flight to another airport during schedule execution, then the airport information for the destination station changes to that of the airport where the flight is to be diverted. It is important to note that this scenario is unlikely, based upon our research scope, because we consider irregular operations for delayed flight schedules only.
    
    \item \textit{Indeterminate aleatoric features}: These are separate data features from flight schedule features that represent disruption types for different functional domains, which occur randomly during schedule execution on day of operation. Examples of indeterminate aleatoric features include delay codes for uncontrollable inclement weather and controllable maintenance inspections. From a disruption management perspective, indeterminate aleatoric features represent triggers for the need of the AOCC to address a specific disruption. As such, indeterminate aleatoric features are data features that can create the most uncertain responses in disruption management initiatives employed by the AOCC during schedule execution. From a statistical perspective, indeterminate aleatoric features are data features that are subject to the most possible uncertainty for the risk of occurrence (or instantiation) of irregular operations during schedule execution, due to inherent randomness of disruption events. For example, inclement weather at a particular airport may require a human specialist in the AOCC to delay the departure of a specific flight at the (origin) airport and reassign some or all of its passengers to another flight with a later departure, while also reallocating the arrival of the original delayed flight to a different gate at the destination airport.    
    
    \item \textit{Epistemic features}: These represent flight schedule features that are determined during the tactical and operational phases of airline planning and can be subject to change during schedule execution on day of operation. Examples of epistemic features include specific departure and arrival times during the day, aircraft type, delay periods, actual turnaround and block time periods. With regards to disruption management, epistemic features represent flight schedule features with considerable amount of alternatives for every time the AOCC initiates a disruption management plan. As such, from a statistical standpoint, epistemic features are flight schedule features that are subject to the most possible uncertainty for the risk of alteration during irregular operations for disruption management, due to lack of knowledge of the exact impact of their alteration. For instance, following a specific disruption like late arrival of flight crew for a scheduled flight, a human specialist in the AOCC may choose to delay the departure of the flight by a specific period of time after the original departure time. However, most times, the human specialist can not guarantee that the decision on applying a particular delay duration after scheduled departure will produce a specific recovery plan, due to the cascading effect of disruptions in large airline networks.  

\end{enumerate}
Fig.~\ref{fig:KAH} shows a generic knowledge abstraction for airline disruption management based upon some specific flight schedule features. The horizontal axis in Fig.~\ref{fig:KAH} represents event abstraction for defining the value of flight operations management based upon the perishable nature of a flight service during schedule execution, while the vertical axis represents uncertainty abstraction for defining the risk of disruption instances and schedule alteration during flight schedule planning.   

\subsection{Feature Transformation}
While the abstraction of raw flight schedule data features provides an excellent avenue for effectively representing latent planning capabilities in airline operations control, the quality of the knowledge extracted from the raw data can be enhanced through transformation to enable discernible representation and interpretation for machine learning algorithms \citep{Liu1998, Kusiak2001}. These algorithms provide efficient means for easily recognizing useful patterns and relationships amongst flight schedule features in a data set. To this end, feature transformation is the process of converting flight schedule and disruption features in raw historical airline  scheduling and operations data into relevant mathematical properties (or functions) that can be readily understood by machine learning algorithms. Every direct flight schedule in the raw data set is defined by forty separate data features (or attributes) that describe different resources, behaviors, and performance indicators that are observable during airline scheduling and disruption management. As such, raw flight schedule features can be separated into four distinct categories namely: geographical features, temporal features, categorical features, and continuous features. 

\begin{itemize}
    \item \textit{Geographical features}: These are flight schedule features which represent resources, behaviors, or performance indicators that require or enable the perception and property of geographic location (or position) during airline disruption management. Examples of geographical features in the raw data set are International Air Transport Association (IATA) codes for departure and arrival airport stations, and the identifier for the origin of the first departure flight of the day. 
    
    \item \textit{Temporal features}: These are flight schedule features that describe and enable the perception and property of time during airline scheduling and disruption management. Temporal features are conceptualized by four different types of time \citep{Shurkhovetskyy2018} namely:
    \begin{enumerate}
        \item \textit{ordinal time}: This represents time points that occur one after another on day of operation. Examples of flight schedule features defined by ordinal time are time-of-day events such as aircraft pushback time, takeoff time, landing time, and aircraft gate-parking time. 
        
        \item \textit{interval time}: This represents time events that are measured on an interval scale with a specific duration (or length). Examples of flight schedule features characterized by interval time include the duration between aircraft pushback and aircraft gate-parking otherwise known as blocktime, the duration for boarding passengers and loading cargo unto an aircraft also known as turnaround, and the duration of any form of delay in airline operations during schedule execution. 
        \item \textit{cyclic time}: This describes cyclic or repeatable processes wherein the application of an ordered relation is inane. Flight date is an example of a flight schedule feature characterized by cyclic time. 
        \item \textit{branching time}: This represents time points that can occur in different branches or alternatives to describe several scenarios or processes. Thus, all temporal flight schedule features in the raw data set are defined by branching time. 
    \end{enumerate}
    \item \textit{Categorical features}: These are flight schedule features that represent fields in the raw data defined by discrete values which belong to a finite set of categories or classes. Categorical features can be text or numeric, and are separated into two classes namely nominal and ordinal, based upon the perception of ordering.
    \begin{enumerate}
        \item \textit{nominal}: Nominal categorical features represent flight schedule features for which there is no concept of ordering among different values of each feature. An example of a nominal categorical feature is A0, which is a binary number (i.e. 0 or 1) indicating whether or not a flight schedule arrives exactly on time.
        \item \textit{ordinal}: Ordinal categorical features represent flight schedule features for which there is a strict adherence to the concept of ordering among different values of each feature. An example of an ordinal categorical feature in flight schedule data is aircraft type, which effectively characterizes the relevance of size and seat capacity for aircraft performance.
    \end{enumerate}
    \item \textit{Continuous features}: These are flight schedule features that represent fields in the raw data, which have infinitely many alternatives between any two values. Examples of continuous features in raw flight schedule data include digital timestamps for different time-of-day events (such as takeoff time) during schedule execution. 
\end{itemize}

\begin{figure}[t!]
\centering
\includegraphics[width=0.99\textwidth]{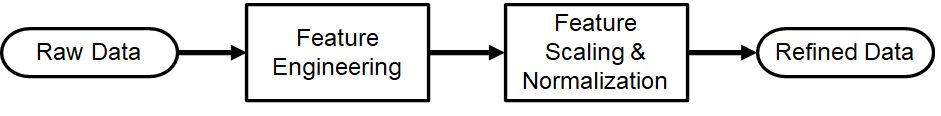}
\caption{Feature transformation process of raw data for airline disruption management}
\label{fig:FT}
\end{figure}

Fig~\ref{fig:FT} reveals a two-layer data transformation process of raw flight schedule data features for airline disruption management. The first layer, known as feature engineering, enables the creation of additional data features from mathematical functions that characterize rudimentary properties of the airline operations control center. Next, these data features are combined with extant low-level data features in the raw data set and normalized by using fundamental statistical parameters in the second layer through a process called feature scaling.

\begin{enumerate}[label=(\roman*)]
    \item \textbf{Feature engineering} represents a \textit{first-degree} transformation of raw flight schedule data that defines the augmentation of properties associated with daily routines of functional roles in the AOCC by using mathematical principles. Thus, flight schedule features representing geographic locations (i.e. geographical features) such as departure and arrival stations are first transformed into spherical directional vectors based upon the longitude and latitude coordinates of their corresponding airport stations, and subsequently transformed into the distance between the departure and arrival airports on an oblate spheroid Earth via the Vincenty geodesic equation \citep{T.Vincenty1975}. Ordinal temporal flight schedule features (such as departure and arrival times) are transformed into two separate periodic (i.e. sine and/or cosine) vectors of different amplitudes, based upon a 24-hr clock period and the percentage of 8-hr work shift completed (at the time of departure or arrival) by human specialists in the AOCC, respectively. The work shift characterization of time-of-day events, via a periodic vector, is intended to capture and represent daily disruption resolution proclivities of human specialists, which can be induced by how much time the specialists have to address a disruption before their work shift is complete. Cyclic temporal flight schedule features defined by Gregorian dates are transformed into four separate periodic vectors, whose periods are based upon the season of the year, month of the year, day of the week, and day of the year respectively.

    Lastly, most categorical flight schedule features defined by texts are transformed into binary numbers by one-hot encoding \citep{Seger2018}, wherein all $n$ feature values are represented as a $n$-dimensional sparse vector with zero entries except for one of the dimensions for which the entry is one. However, categorical flight schedule feature values for aircraft type, which are defined by aircraft model codes, are transformed into discrete numbers based upon the total number of available seats in the aircraft. A secondary objective of feature engineering is to provide a precursor for continuous feature representation by ensuring that all data features are in numeric form. As such, feature engineering may not be applicable to flight schedule features that are already in continuous form in a raw data set.
    
    \item \textbf{Feature Scaling} represents a \textit{second-degree} transformation of raw numeric data features that include additional features created from feature engineering, such that a uniform statistical grounding basis is used to transform values for all data fields (i.e. flight schedule features) in the raw data set into bounded continuous values that describe a differentiable function. We explore three different approaches for enabling feature normalization \citep{Pedregosa2011} namely: \textit{Standard scaling}, \textit{Range scaling}, and \textit{Power scaling}. 
    \begin{enumerate}
        \item \textit{Standard scaling}: Standard scaling or standardization normalizes values for each flight schedule feature in the data set by removing the mean of the values and scaling to a unit variance, thus resulting in a standard score for each value. The standard score, $z_{i}$, of an arbitrary sample (i.e data feature value), $x_{i}$, in the data set is calculated as follows: 
        \begin{align}
            z_{i} = \frac{(x_{i}-u)}{s}
        \end{align}
        where $u$ and $s$ represent the mean and standard deviation, respectively, of all values for each flight schedule feature in the data set. As such, standardization provides a platform to ensure that each flight schedule feature in the data set follows a Gaussian distribution with zero mean and a variance of one.    
        
        \item \textit{Range scaling}: Range scaling, otherwise known as min-max normalization, transforms each flight schedule feature in the data set by scaling the values of each feature by the difference between the maximum value and the minimum value (i.e. range). This results in adjusted values of a range (or distance) between zero and one for each flight schedule feature. The adjusted (range-scaled) value, $y_{i}$, of a characteristic flight schedule feature in the data set is calculated as follows:  
        \begin{align}
            y_{i} = \frac{x_{i} -\min(X)}{\max{(X)} - \min{(X)}}
        \end{align}
        where $x_{i}$ and $X$ represent an original value and set of all original values, respectively, for a flight schedule feature. 
        \item \textit{Power scaling}: Power scaling involves adapting a family of parametric and monotonic transformations to convert flight schedule data values from any distribution to the closest possible representation of Gaussian distribution, so as to reduce variance and skewness in data. An appropriate power transformation of flight schedule and disruption features is the Yeo-Johnson transform \citep{Weisberg2001}, because it can be applied to all forms of numeric data just like standard and range scaling transforms. The Yeo-Johnson transform is given by: 
        \begin{align}   
            x_{i}^{(\lambda)} = 
                \begin{cases}
                    [(x_{i}+1)^{\lambda} -1]/\lambda &\quad\text{if} \quad \lambda \neq 0, x_{i} \geq 0, \\
                    \ln{(x_{i}+1)} &\quad\text{ if} \quad \lambda = 0, x_{i} \geq 0, \\
                    -[(-x_{i}+1)^{2 -\lambda} -1]/(2-\lambda) &\quad\text{ if} \quad \lambda \neq 2, x_{i} < 0,\\
                    -\ln{(-x_{i}+1)} &\quad\text{ if} \quad \lambda = 2, x_{i} < 0 \\ 
            \end{cases}
        \end{align}
    where $x_{i}$ and $\lambda$ represent an original data value and an arbitrary parameter that is determined through maximum likelihood estimation \citep{Glas2017}, respectively. 
    \end{enumerate}
\end{enumerate}

Completion of the feature transformation process shown in Fig.~\ref{fig:FT} results in a refined, continuous data set that can be readily comprehensible by suitable machine learning estimators.  

\subsection{Dimensionality Reduction}\label{dim_red}
The efficacy of constructing and applying relevant machine learning algorithms for identifying and acknowledging high-level properties from data features (such as flight schedule and disruption data features) is dependent on the form in which the data values are presented. To this end, feature transformation has a strong propensity to increase the number of elements in the flight schedule and disruption feature space that constitutes the problem dimensions for airline disruption management. As such, the intrinsic dimensionality of the refined data appropriated for airline disruption management is defined by the least number of flight schedule and disruption features required to delineate observed behavioral properties from AOCC routines. Hence, dimensionality reduction is the process of mitigating the curse of dimensionality \citep{Verleysen2005} and other unwanted properties of high-dimensional feature space through classification, visualization, and compression of high dimensional data obtained as a result of feature transformation \citep{VanDerMaaten2008}. In essence, dimensionality reduction aims to provide a rudimentary means to attain and observe the latent feature space of a refined data set for airline disruption management.     

From a mathematical perspective, we assume that the refined flight schedule and disruption data set is represented in a $n \times m$ matrix $\textbf{X}$, which consists of $n$ feature vectors $\textbf{x}_{i} (i \in \{1, 2, ... , n\})$ with dimensionality $m$. Furthermore, we assume that the refined data set has an intrinsic dimensionality $d$, such that $d < m$ and often $d || m$. The intrinsic dimensionality property refers to points in the refined data set $\textbf{X}$, which lie near a manifold with dimensionality $d$ that is embedded in the $m$-dimensional feature space. To that effect, dimensionality reduction techniques transmute the refined flight schedule and disruption data set $\textbf{X}$ with dimensionality $m$ into a new data set $\textbf{Y}$ with dimensionality $d$, while maintaining the geometry of the refined data set $\textbf{X}$ as much as possible. Typically, the intrinsic dimensionality $d$ and the geometry of the manifold of the new data set $\textbf{Y}$ are unknown, and as such, most dimensionality reduction techniques require that certain assumptions about the properties (like intrinsic dimensionality) of the refined data set be made a priori. For the remainder of this section, we denote a high dimensional data instance for flight schedule and disruption (i.e. datapoint) by $\textbf{x}_{i}$, such that $\textbf{x}_{i}$ is the $i^{th}$ row of the refined $m$-dimensional data set $\textbf{X}$. In complement, the low-dimensional equivalent of $\textbf{x}_{i}$ is expressed by $\textbf{y}_{i}$, where $\textbf{y}_{i}$ is the $i^{th}$ row of the new $d$-dimensional matrix $\textbf{Y}$.

To demonstrate the usefulness of dimensionality reduction on refined flight schedule and operations data, we investigate two separate techniques that employ linear and nonlinear principles nicknamed \textit{PCA} and \textit{t-SNE}, respectively, by utilizing delayed flight schedule and disruption instances for the Weather functional domain in SWA-NOC between September 2016 and September 2017. It is important to note that the flight schedule data for the Weather functional domain constitutes a subset (with 12,659 delayed flight schedule instances) of the full refined data set. For validation, we adopt min-max normalization for scaling all feature values in the refined data set because both dimensionality reduction techniques strongly depend on Euclidean distances between refined high-dimensional datapoints $\textbf{x}_{i}$ and $\textbf{x}_{j}$ to obtain and simplify the gradient of their respective cost functions \citep{VanDerMaaten2008, VanDerMaaten2009}.     

\subsubsection{Principal Component Analysis (PCA)}
Principal component analysis or \textit{PCA} is a standard non-parametric tool in modern data analysis used for extracting relevant information from large and confusing data sets~\citep{Shlens2014}. \textit{PCA} is also a full spectral linear technique for dimensionality reduction that embeds data into a linear subspace of lower dimensionality. In the lower dimension, the refined variables (or data features) in the data set are transformed into linear combinations of the data features, which are called principal components. With minimal effort, \textit{PCA} provides a schema for reducing a fairly complex data set to a lower dimension in order to show simplified structures that often define it, by revealing as much of the variance in the data as possible. As such, the first and second principal components are the orthogonal linear combinations of the refined data features that have the largest and second-largest possible variance (or inertia), respectively, in the refined data set. In mathematical terms, \textit{PCA} aims to find a linear mapping $\textbf{X}$ that maximizes the variance (or minimizes the reconstruction error) defined by $\mathrm{trace}(\textbf{S}^{T}\mathrm{cov}(\textbf{X})\textbf{S}))$, wherein $\mathrm{cov}(X)$ is the sample covariance matrix of the refined data \citep{Wold1987a}. Thus, the linear mapping created by the $d$ principal components (or principal eigenvectors) are solutions to the eigenproblem defined as follows:
    
\begin{align}
    \mathrm{cov}(\textbf{X}) = \lambda(\textbf{S})
\end{align}
    
\begin{figure}[h!]
    \centering
    \includegraphics[width=0.99\textwidth]{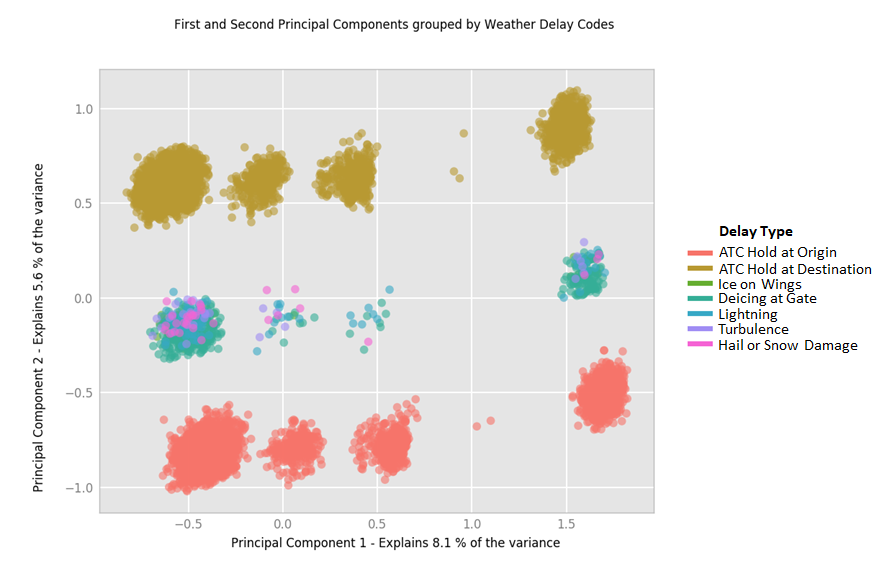}
    \caption{Principal component analysis of indeterminate aleatoric features for Weather domain}
    \label{fig:PCA}
\end{figure}
    
The lower dimensional representation of the refined flight schedule and disruption feature instances, defined by $\textbf{y}_{i}$ of $\textbf{x}_{i}$ datapoints, are computed by mapping them onto a linear basis $\textbf{Y} = \textbf{X}\textbf{S}$ that solves the eigenproblem for the $d$ principal eigenvalues defined by $\lambda$, via the scikit-learn software \citep{Pedregosa2011}.
    
Fig.~\ref{fig:PCA} presents a visualization of the analysis for the first two principal components describing indeterminate aleatoric features that represent delays for the Weather functional domain in SWA-NOC. The first and second principal components represent orthogonal linear combinations of the refined flight schedule features that account for 8.1\% and 5.6\%, respectively, of the variance of indeterminate aleatoric features related to weather delays in the data set. Fig.~\ref{fig:PCA} reveals that there are four major types of weather-related delays (\textit{ATC Hold at Origin}, \textit{ATC Hold at Destination}, \textit{Deicing at Gate}, and \textit{Hail or Snow Damage}) in the data set. \textit{ATC Hold at Origin} and \textit{ATC Hold at Destination} represent weather delays due to gate hold from air traffic control (ATC) at departure and arrival stations respectively. \textit{Deicing at Gate} and \textit{Hail or Snow Damage} represent weather delays due to deicing at the gate, and aircraft swap due to hail or snow damage respectively. Fig.~\ref{fig:PCA} shows that the data set is divided into four separate clusters of the same delay type along the axis of the first principal component. This reveals that the axis of the first principal component (horizontal) represents linear combinations of flight schedule features that capture the seasonal behavior of weather-related delays, as each data cluster describes each weather season over the one-year data-collation period. Furthermore, the data set is divided into two polarizing (\textit{ATC Hold at Origin} and \textit{ATC Hold at Destination}) and two overlapping (\textit{Deicing at Gate} and \textit{Hail and Snow Damage}) clusters along the second principal component (vertical) axis. This shows that the axis of the second principal component represents linear combinations of flight schedule features that capture the difference in the types of indeterminate aleatoric features for weather-related delays in the refined data set. 
    
The patterns and information gleaned from the results and observations from the \textit{PCA} method can be appropriated for informing model development for airline disruption management. For instance, the seasonal relationship among the four predominant types of weather-related delays observed in Fig.~\ref{fig:PCA} can be quantified via a linear combination of flight schedule features, which suggests that decision-making by human specialists in the AOCC is sensitive to weather seasons. In addition, the overlapping effect observed between \textit{Deicing at Gate} and \textit{Hail and Snow Damage} in Fig.~\ref{fig:PCA} is representative of similarities in the type of disruption resolutions used for weather-related delays during the winter season, thus bolstering the significance of the effect of seasonal properties on decision-making by human specialists in the AOCC.
    
\subsubsection{t-distributed Stochastic Neighborhood Embedding (t-SNE)}
t-distributed Stochastic neighborhood embedding or \textit{t-SNE} represents a recent advancement in clutering and visualization for dimensionality reduction that provides a nonlinear platform for transforming the Euclidean distances between refined values (i.e. datapoints) of flight schedule and disruption features into conditional probabilities that define similarities. As such, the similarity of a datapoint $x_{j}$ to another datapoint $x_{i}$ is the conditional probability ($p_{j|i}$) that $x_{i}$ will select $x_{j}$ as its neighbor if neighbors are selected in proportion to their probability density under a Student-t distribution with one degree of freedom (i.e. Cauchy distribution) centered about $x_{i}$ \citep{VanDerMaaten2008, Maaten2014}. Thus, $p_{j|i}$ remains comparatively high for datapoints in close proximity and insignificant for datapoints that are substantially separated. Mathematically, the objective of \textit{t-SNE} is to minimize the Kullback-Leibler divergence \citep{Perez-Cruz2008, Joyce2011} between a joint probability distribution defined by \textit{P} in the high-dimensional feature space and a joint probability distribution defined by \textit{Q} in the low-dimensional feature space. Hence, the Kullback-Leibler divergence represents the cost function of the following optimization problem, which is solved via the scikit-learn software \citep{Pedregosa2011}:
    
    \begin{figure}[h!]
        \centering
        \includegraphics[width=0.99\textwidth]{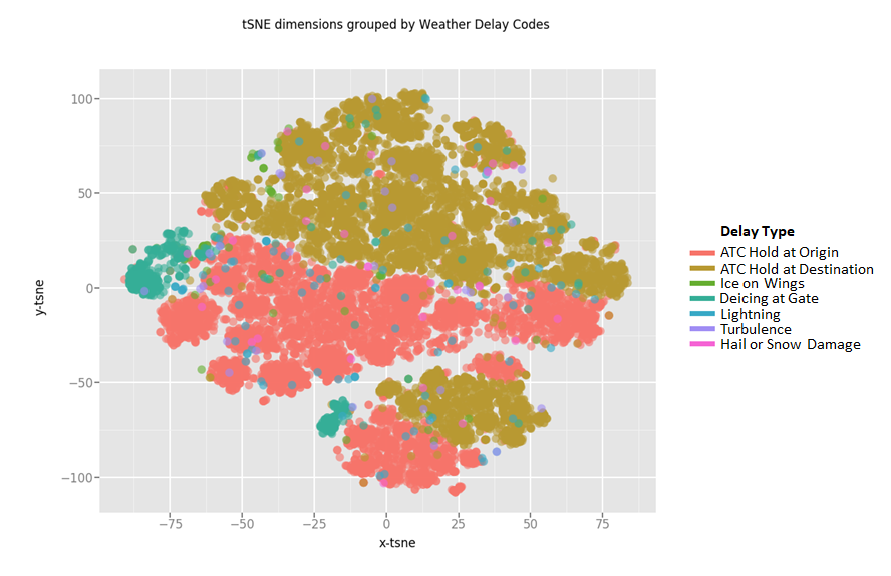}
        \caption{t-Distributed Stochastic Neighborhood Embedding analysis of indeterminate aleatoric features for Weather domain}
        \label{fig:t-SNE}
    \end{figure}
    
    \begin{equation}
        \begin{aligned}
        \min \quad & \operatorname{KL}(\textit{P}||\textit{Q}) = \sum_{i}\sum_{j}{p_{ij}\log{\frac{p_{ij}}{q_{ij}}}}\\
        \textrm{s.t.} \quad & p_{ij} = \frac{\exp{(-||{x_{i} - x_{j}}||^{2}/2\sigma^{2})}}{\sum_{k\neq l}{\exp{(-||{x_{i} - x_{j}}||^{2}/2\sigma^{2})}}}\\
        & q_{ij} = \frac{(1 + ||{y_{i} - y_{j}}||^{2})^{-1}}{\sum_{k\neq l}{({1 + ||{y_{k} - y_{l}}||^{2})^{-1}}}}\\ 
        \end{aligned}
    \end{equation}
    where $p_{ii}$ and $q_{ii}$ are set to zero, and $p_{ij} = p_{ji}$ and $q_{ij} = q_{ji}$ for all ${i, j}$. 
    
    Fig.~\ref{fig:t-SNE} presents a strictly visual perception of the \textit{t-SNE} nonlinear projection of the two-dimensional space for flight schedule features, which describes indeterminate aleatoric features that represent delays for the Weather functional domain in SWA-NOC. Similar to observations from \textit{PCA}, the red and gold clusters in Fig.~\ref{fig:t-SNE} reveal that weather-related delays due to \textit{ATC Hold at Origin} and \textit{ATC Hold at Destination} are the most prominent and oppositely related, based upon the symmetry observed from the small and large blobs of red and gold clusters. As such, the polarizing effect observed between \textit{ATC Hold at Origin} and \textit{ATC Hold at Destination} in Figs.~\ref{fig:PCA} and \ref{fig:t-SNE} can be attributed to the importance of the geographical location (i.e departure or arrival stations) on how weather-related disruption resolutions are applied in the AOCC. As such, flight schedule features associated with geographical location are relevant for creating robust models for airline disruption management.
    
\subsection{Feature Selection}
Although dimensionality reduction techniques provide an effective means to readily (i.e. visually) discern high-level patterns and properties associated with a data set, they are ineffectual in revealing detailed information on the specific importance of flight schedule and disruption features in a data set and their corresponding relationships \citep{Koller1996}. To this end, feature selection presents simple fundamental methods for efficiently selecting and investigating pertinent associations among data features in a refined data set, which can provide insightful knowledge (or a priori information) for developing useful data-driven models for robust airline disruption management. In essence, feature selection methods aim to proactively enhance model prediction performances by increasing generalization (i.e. minimize data overfitting) and decreasing model runtimes \citep{Moran2019}. There are three major categories of feature selection methods namely: wrapper, filter, and embedded methods. 

Wrapper methods involve algorithms that search the feature space for plausible subsets of features by assessing each subset after running a specific model. Typically, the model is validated on a test data set to estimate the model's error rate, before a score is registered for each feature subset and the feature subset with the best score is ultimately selected. Unlike computationally intensive wrapper methods, filter methods do not consider a model when searching the feature space for relevant subsets of the feature space, and rely on general statistical measures such as Pearson correlation coefficient \citep{Benesty2009} and mutual information \citep{Jiang2010}. In this manner, filter methods are somewhat analogous to dimensionality reduction techniques, such that they are not customized to a particular type of predictive model and consume significantly less computational resources than wrapper methods. Embedded methods involve feature selection methods that are entrenched in a specific learning algorithm that performs classification (or regression) and feature selection concurrently. As such, embedded methods deliver the advantages of both wrapper and filter methods with medium computational expense. 

To demonstrate the relevance of feature selection on refined flight scheduling and operations data, we apply two specific types of feature selection that belong to the filter and embedded categories, respectively, to identify flight schedule features that are pertinent for disruption management during turnaround. Turnaround is an airline process (or time period) primarily representative of loading, unloading and occasional servicing of aircraft, and is crucial for minimizing overall flight schedule delays. In addition to reducing overall flight delay, most airlines typically aim to expedite the turnaround process as much as possible in order to avoid causing discomfort to passengers, stemming from long waits in the aircraft on the ground, thus invariably minimizing loss of passenger goodwill.

In that regard, the filter method that we apply is defined by mutual information and the embedded method is defined by a Gaussian process, such that actual turnaround duration is set as the target flight schedule feature. We do not consider wrapper methods in our discussion because of the significant computational expense required as compared to filter and embedded methods. Similar to the dimensionality reduction analysis discussed in Section~\ref{dim_red}, we utilize delayed flight schedule and disruption instances for the Weather functional domain in SWA-NOC between September 2016 and September 2017 for our analysis. For validation, we adopt standardization for the \textit{second-degree} transformation (i.e. scaling) of all feature (i.e label and target) values in the refined data set, because the algorithms for both filter and embedded methods perform best with a zero-mean Gaussian distribution as prior instantiation for each feature space in the refined data set \citep{C.E.Rasmussen2006, Jiang2010, Ross2014}.

\subsubsection{Mutual Information Regression (MIR)}
Mutual information is a non-negative measure from information theory that provides an excellent statistic for quantifying the degree of relatedness among flight schedule and disruption features in a refined data set. In that regard, mutual information is closely related to the entropy of a flight schedule feature based upon observing another flight schedule feature in a refined data set \citep{Kraskov2004}. In addition to the ability to readily identify relationships amongst data features, mutual information provides a fundamental metric for straightforward interpretation of the relationships among data features as shared information (i.e. shannons or bits) between data features. To that effect, mutual information is insensitive to the number of instances in a data set \citep{Ross2014}. 
    
 Mathematically, mutual information, $I$, is expressed as:
    \begin{align}
        I(X;Y) = \int_{X}\int_{Y} p(x,y)\log{\frac{p(x,y)}{p(x)p(y)}}dxdy
    \end{align}
    
    where $p(x,y)$ is the joint probability density function of $X$ and $Y$, and $p(x)$ and $p(y)$ are the marginal probability density functions of $X$ and $Y$ respectively. Thus, for feature selection, the objective of \textit{MIR} is to maximize the mutual information between a subset of flight schedule features defined by $\textbf{X}_{s}$ and a target flight schedule feature defined by $y$ as 
    represented by the following optimization problem:
    
    \begin{figure}[t!]
        \centering
        \includegraphics[width=0.99\textwidth]{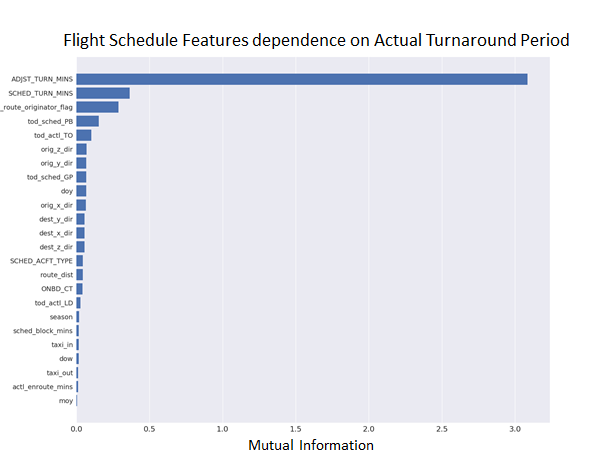}
        \caption{Mutual information of flight schedule data features for Weather Domain with respect to actual turnaround duration}
        \label{fig:MIR}
    \end{figure}
    
    \begin{equation}
        \begin{aligned}
            s^{*} =\argmax_{s}
            I(\textbf{X}_{s};y) \quad \textrm{s.t.} \quad & |s| = k \\
        \end{aligned}
    \end{equation}

where $k$ is the number of features that are to be selected. A non-parametric regression algorithm based upon entropy estimation from k-nearest neighbor distances is used to solve the NP-hard optimization problem via the scikit-learn software \citep{Pedregosa2011}, for a set of possible combinations of data features that increases exponentially \citep{Kraskov2004, Gao2015}. 

Fig.~\ref{fig:MIR} shows the mutual dependence, in decreasing order, of flight schedule features (i.e. determinate aleatoric and epistemic features) on actual turnaround period for instances of flight delay from the Weather functional domain. Fig.~\ref{fig:MIR} reveals that turnaround duration adjusted during schedule execution has the highest mutual information of over 3, thus implying that it has the strongest mutual dependency on the decison-making for estimating actual turnaround duration during disruption management. In addition, turnaround period estimated prior to schedule execution and a flight’s capacity to be a route originator (i.e. first departure flight of the day) have the second and third highest mutual information of 0.4 and 0.3 respectively, thereby revealing a weak mutual dependency on the estimation of actual turnaround schedule for managing weather-related delay of flight schedule. Month of the year (\textit{moy}) can not be selected as a significant predictor for estimating actual turnaround duration because of zero mutual dependency as shown in Fig.~\ref{fig:MIR}. 

\subsubsection{Gaussian Process Regression (GPR)}
A Gaussian process is a stochastic process (i.e. random variables indexed by time or space) where a finite collection of random variables have a multivariate normal distribution \citep{C.E.Rasmussen2006}. As such, Gaussian Process Regression or \textit{GPR} is the inference of continuous feature values with a Gaussian process (or distribution) prior, such that the marginal likelihood of the data is maximized \citep{C.E.Rasmussen2006}. Converse to \textit{MIR}, which is a model-free method for dimensionality reduction and feature selection, \textit{GPR} is an embedded method that offers nonlinear and non-parametric regression properties that enable the natural decomposition of flight schedule features in an airline data set for simultaneously attaining high fidelity dimensionality reduction and feature selection. \textit{GPR} provides an appropriate medium to obtain the sensitivity and importance of flight schedule and disruption features necessary for informing the development of appropriate models for airline disruption management.  
 
Following the nomenclature in \cite{C.E.Rasmussen2006}, consider a training data set defined by $\mathcal{D}$ with $n$ observations, such that $\mathcal{D} = \{(\textbf{x}_{i}, y_{i}) | i = 1,...,n\}$, where $\textbf{x}$ represents an input vector of flight schedule features of dimension $D$ and $y$ is a scalar target or output feature. As such, the column input vector of flight schedule features for all $n$ instances are collected in a matrix $\textbf{X}$ of size $D \times n$, and scalar outcomes of the target feature are collected in the vector $\textbf{y}$ to yield $\mathcal{D}=(\textbf{X},\textbf{y})$. \textit{GPR} assumes that outcomes of a target feature defined by $\textbf{y}$ are noisy observations of an unknown function of the input flight schedule features $\textbf{x}$ such that $y = f(\textbf{x}) + \epsilon$, where $\epsilon$ represents independent and identically distributed zero mean Gaussian random variables with unknown variance $\sigma_{n}^{2}$. Thus, a Gaussian process prior is placed over all values of $f(\textbf{x})$ before $y$ is observed, such that $f(\textbf{x})|\theta \sim \mathcal{GP}(m(\textbf{x};\theta), K(\textbf{x},\textbf{x};\theta))$ and $m(\textbf{x})$, $K(\textbf{x},\textbf{x})$ and $\theta$ represent a mean function, a covariance function, and all hyperparameters that influence the mean and covariance functions, respectively. A combination of Bayes' rule and Gaussian identities \citep{C.E.Rasmussen2006} provides a method to retrieve the posterior distribution of the values of $f(\textbf{x})$ after observing $y$, such that $\sigma_{n}^2 \in \theta$ albeit the mean and covariance functions are independent of $\theta$ \citep{Lee2020}. As such, the posterior distribution is defined by the analytical solution for $\textbf{f}$:

\begin{equation}\label{gpr_eqn1}
    \begin{aligned} 
        \textbf{f}|\textbf{y},\textbf{X} \sim \mathcal{N}(\mu + KK_{n}^{-1}(\textbf{y}-\mu), K - KK_{n}^{-1}K),
    \end{aligned}
\end{equation}

where $\mu$ is the prior mean of $\textbf{f}$, $K$ represents the prior covariance function evaluated at $\textbf{X}$, and $K_{n}^{-1} = (K+\sigma_{n}^{2}I)^{-1}$. In order to obtain predictions of target outcomes for new test cases represented by a collection of input flight schedule features $\textbf{X}^{*}$, $\textbf{f}^{*}$ is defined as: 
\begin{equation}\label{gpr_eqn2}
    \begin{aligned} 
        \textbf{f}^{*}|\textbf{y},\textbf{X},\textbf{X}^{*} \sim \mathcal{N}(\mu^{*} + K(\textbf{X}^{*},\textbf{X})K_{n}^{-1}(\textbf{y}-\mu), \\ K(\textbf{X}^{*})-K(\textbf{X}^{*},\textbf{X})K_n^{-1}K(\textbf{X},\textbf{X}^{*})),
    \end{aligned}
\end{equation}
 where $\mu^{*}$ is the prior mean function evaluated at $\textbf{X}^{*}$. 

Computationally, \textit{GPR} estimates the maximum marginal log likelihood of the distribution for the target flight schedule feature in a training data set, such that hyperparameters (or lengthscales) associated with all input flight schedule features that define different drivers of disruption management by the AOCC are optimized with respect to a certain kernel (covariance) function.

\begin{figure}[h]
	\centering
	\includegraphics[width=1\textwidth]{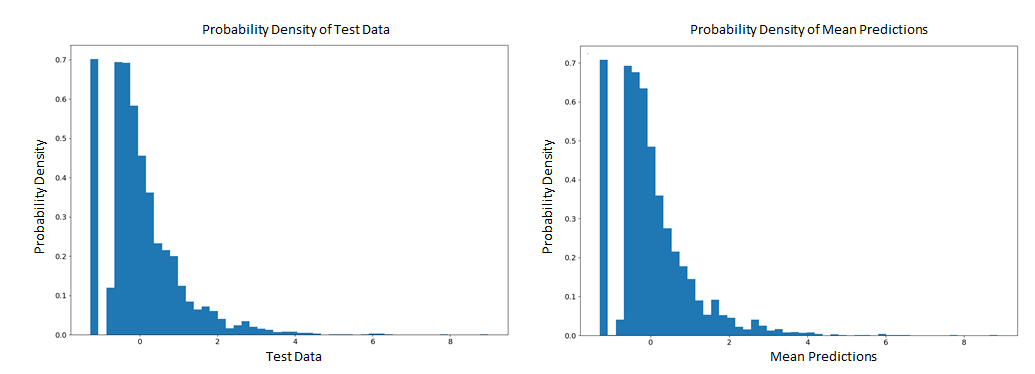}
	\caption{Probability densities of test turnaround duration and mean predictions of turnaround duration for delayed flight schedule instances in Weather domain}
	\label{fig:data_distributions}
\end{figure}
\begin{figure}[h!]
	\centering
	\includegraphics[width=1\textwidth]{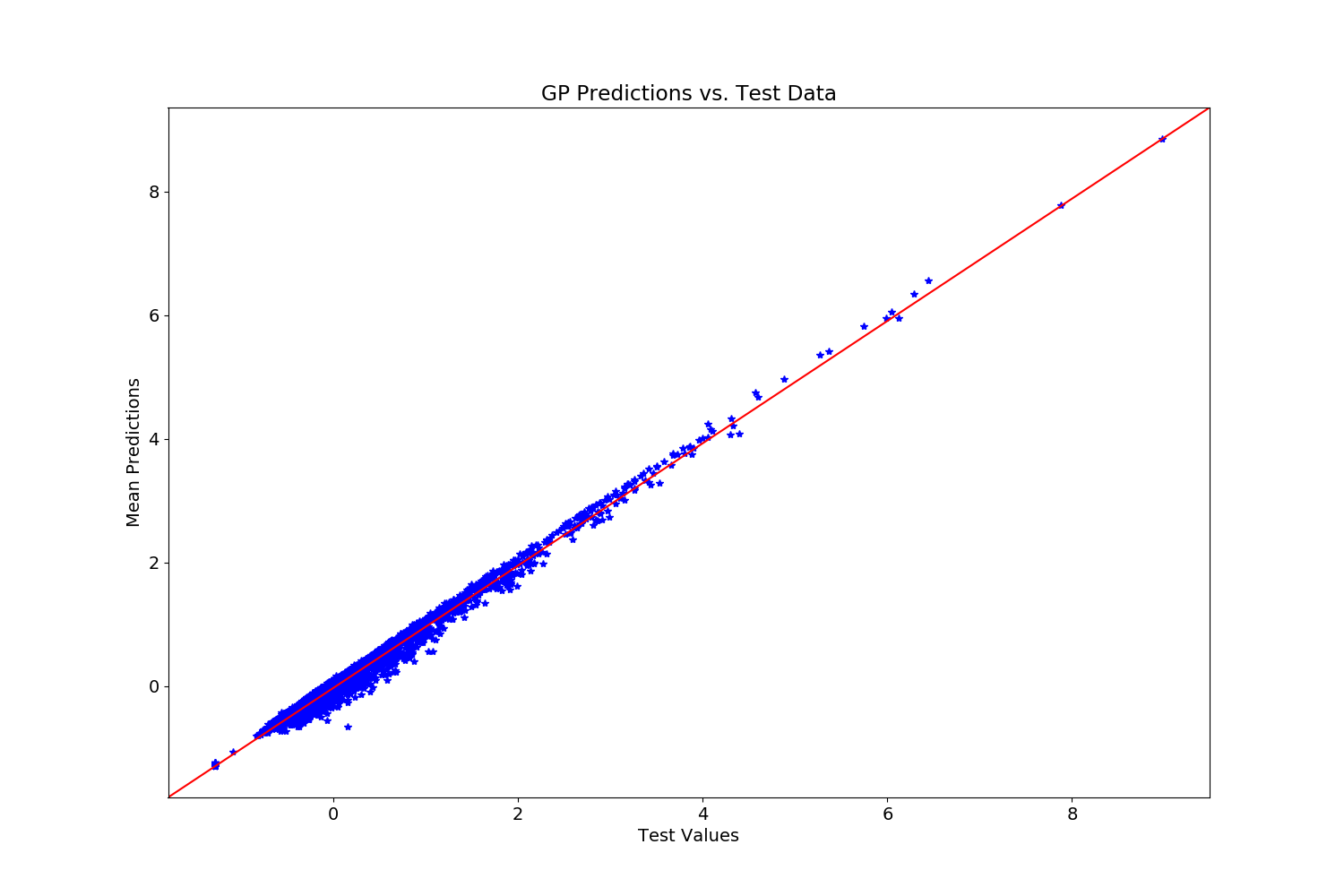}
	\caption{Mean predicted turnaround duration data vs. test turnaround duration data}
	\label{fig:GP_Predictions}
\end{figure}

As previously mentioned, a subset of the data set defined only by flight schedules delayed by weather incidents is used for the \textit{GPR} demonstration. This subset of data is split into two separate sets of training and test (new) data respectively, such that the training data (70\% of the data subset randomly selected) is used to fit the \textit{GPR} model for actual turnaround duration and the test data (i.e. remaining 30\% of the data subset that is unseen) is used to validate the model by verifying that the test data is consistent with mean predictions from the model. Plotting the probability density function of the test data, revealed a lognormal distribution of the actual turnaround duration in the data set, as evidenced by Fig.~\ref{fig:data_distributions}. Hence, the Matern32 kernel function, which is a combination of Gamma and Bessel functions correlated by an hyperparameter of 3/2, is selected to fit the \textit{GPR} model by means of a Gaussian process software named GPy \citep{sheffield2014}.

\begin{figure}[h!]
	\centering
	\includegraphics[width=1\textwidth]{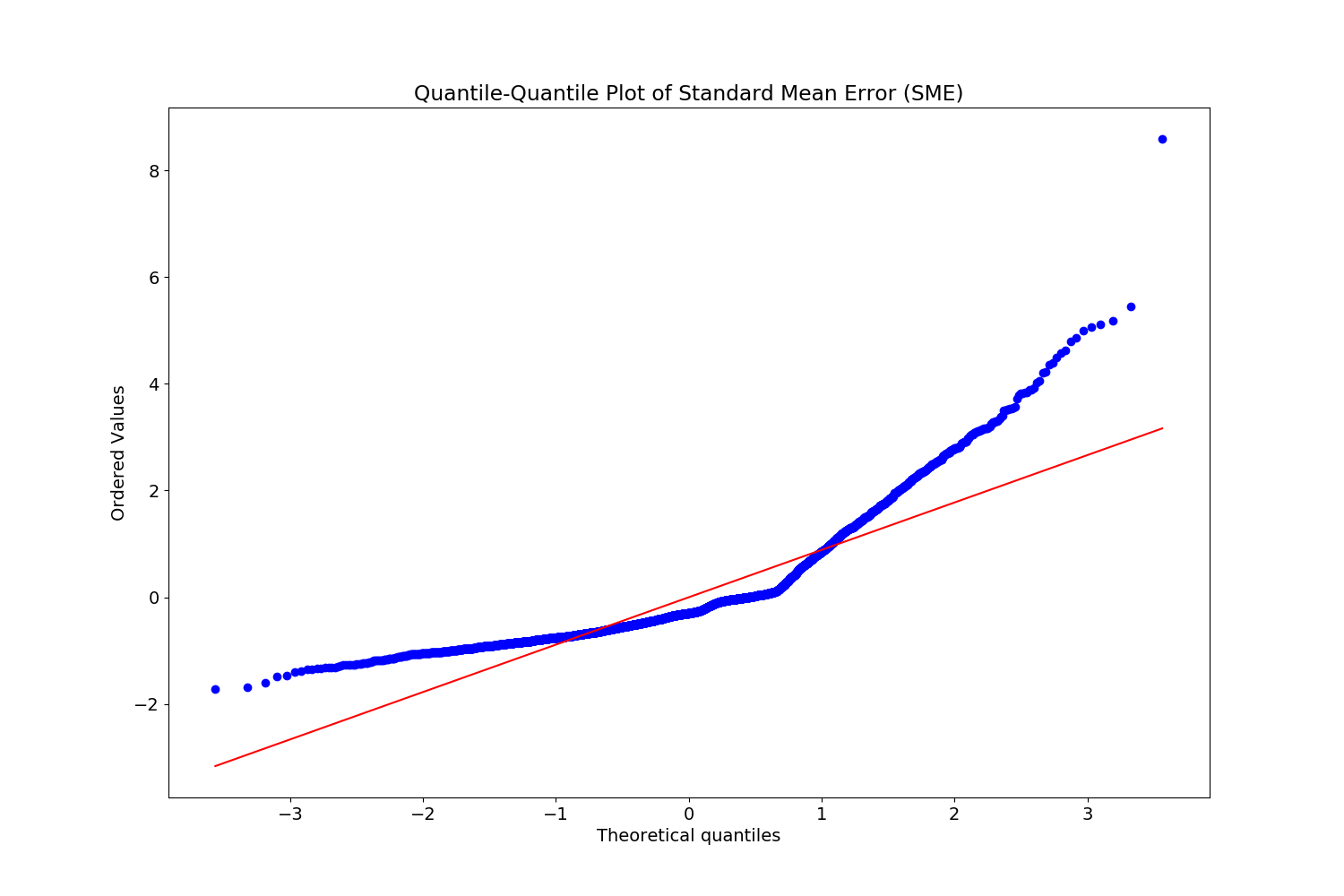}
	\caption{Quantile-Quantile plot of standard mean error between predicted and test data for actual turnaround duration} 
	\label{fig:qqplot}
\end{figure}

\begin{table}[!]
\caption{Lengthscales of refined data features for predicting actual turnaround duration through Gaussian process regression}
\begin{center}
\begin{tabular}{*3c}
    \hline
	\textbf{GPR.Mat32.lengthscale} & \textbf{Feature Class} & \textbf{Feature Name} \\ \hline
	1738.68 & Determinate Aleatoric &\textit{sin\_date} \\ \hline
	1650.37 & Determinate Aleatoric &\textit{cos\_date} \\ \hline
	1857.16 & Determinate Aleatoric &\textit{orig\_x\_dir} \\ \hline
	1808.49 & Determinate Aleatoric &\textit{orig\_y\_dir} \\ \hline
	1652.50 & Determinate Aleatoric &\textit{orig\_z\_dir} \\ \hline
	1750.13 & Determinate Aleatoric &\textit{ONBD\_CT} \\ \hline
	1193.69 & Determinate Aleatoric &\textit{SCHED\_TURN\_MINS} \\ \hline
	44.77 & Epistemic&\textit{ADJST\_TURN\_MINS} \\ \hline
	1367.11 & Determinate Aleatoric&\textit{schd\_acft\_type} \\ \hline
	1367.11 & Epistemic&\textit{actl\_acft\_type} \\ \hline
	1406.15 & Epistemic&\textit{SWAP\_FLT\_FLAG} \\ \hline
	975.98 & Indeterminate Aleatoric&\textit{ATC Hold at Origin} \\ \hline
	972.46 & Indeterminate Aleatoric&\textit{ATC Hold at Destination} \\ \hline
	1.00 & Indeterminate Aleatoric&\textit{Deicing at Gate} \\ \hline
	156.75 & Indeterminate Aleatoric&\textit{Ice on Wings} \\ \hline
	1.00 & Indeterminate Aleatoric&\textit{Lightning Strike} \\ \hline
	1.00 & Indeterminate Aleatoric&\textit{Turbulence} \\ \hline
	1.00 & Indeterminate Aleatoric&\textit{Hail or Snow Damage} \\ \hline
\end{tabular}
\end{center}
\label{tab:lengthscales}
\end{table}

Fig.~\ref{fig:GP_Predictions} shows the plot of the mean predictions of the actual turnaround duration from the \textit{GPR} model versus the actual turnaround duration from the test data, for which the turnaround duration values in both axes are scaled to a unit variance from the mean of the data values. The red diagonal line in Fig.~\ref{fig:GP_Predictions} represents the 45-degree line, while each blue star represents a coordinate of the mean \textit{GPR} prediction and test data describing actual turnaround duration for each datapoint (i.e. instance of weather-delayed flight schedule). Fig.~\ref{fig:GP_Predictions} shows that the coordinates for the datapoints follow the trend of red diagonal line almost perfectly (root mean square error of 9\%), which indicates that the \textit{GPR} model is able to effectively predict ``unknown" actual turnaround duration. Each coordinate that falls on the red line implies an exact prediction of the test data by the \textit{GPR} model, and as such, Fig.~\ref{fig:GP_Predictions} shows that the \textit{GPR} model perfectly predicts actual turnaround periods that lie over six standard deviations away from the mean. 

Table~\ref{tab:lengthscales} shows the values of the optimized hyperparameters (i.e. lengthscales) of refined flight schedule and disruption features for estimating actual turnaround time. Lower values in Table~\ref{tab:lengthscales} indicate higher importance of features for predicting actual turnaround period. Similar to the result from \textit{MIR}, turnaround duration adjusted during schedule execution (i.e. \textit{ADJST\_TURN\_MINS}) is the most significant epistemic flight schedule feature for predicting actual turnaround duration, as indicated by its low lengthscale value of approximately 45. Of all the aleatoric features (determinate and indeterminate), disruption features representing deicing at the gate, lightning strike, turbulence, and hail and snow damage (all with lengthscale values of 1) are the most significant for accurately estimating actual turnaround period during schedule execution.       

Fig.~\ref{fig:qqplot} shows the quantile-quantile (QQ) plot of the standard mean error (SME) between the mean predictions from the \textit{GPR} model and the test data for actual turnaround duration. The straight red line in Fig.~\ref{fig:qqplot} represents the trend line for a standard normal distribution, and as such, the bilinear trend for standard mean error (portrayed by the blue dots) in Fig.~\ref{fig:qqplot} confirms that the distribution of actual turnaround period for weather-delayed flight schedules is lognormal. Categorically, the overlapping trend between the 45-degree red line and the spread of the coordinates in Fig.~\ref{fig:GP_Predictions}, coupled with the lognormal distribution trend noted from the QQ plot in Fig.~\ref{fig:qqplot}, validates that the turnaround process for weather-delayed flight schedules is indeed a Gaussian process. To that effect, the relationship between the data features (shown in Table~\ref{tab:lengthscales}) and the actual turnaround duration during schedule execution (i.e. target feature) can be described by a Matern32 covariance function.

\section{Conclusion} \label{conclusion}
This paper provided macroscopic and microscopic analysis of the historical airline scheduling and operations data necessary for addressing disruption management at a major airline in the United States. Through macroscopic analysis, we identified that over $94\%$ of irregular operations over a one-year period occurred due to different forms of flight schedule delays. To that end, we investigated crucial drivers and properties for effectively managing flight schedule delays through microscopic analysis of weather-delayed flight schedule data, which also demonstrated a toolbox for applying appropriate machine learning techniques to enable data-driven schedule recovery during airline disruption management. 

Furthermore, machine learning techniques such as feature transformation and dimensionality reduction provided excellent platforms for verifying empirical processes and validating domain knowledge for airline disruption management. The patterns and information gleaned from the results and observations from visual techniques (such as \textit{PCA} and \textit{t-SNE}) identified the viability of seasonal (temporal) and geographical features in the data set as suitable predictors for artificial intelligence (AI) models that describe disruption management tendencies of the airline. The results obtained from qualitative methods (such as \textit{MIR} and \textit{GPR}) established the importance of the predictors and the underlying processes that define their combination for high fidelity AI models. As such, the findings from our research reveal that the statistics of machine learning can be used to characterize and assess the functional parts of the AOCC for intelligent decision support systems. 

While this paper provides high fidelity exploratory data analysis for airline disruption management, there exist a few areas for further research. First, the data utilized for the analysis was provided by a single U.S. airline that operates a point-to-point route network and may not completely represent the disruption management propensities of other airlines in the United States. As such, there is a need for exploratory analysis on a data set from an airline that operates a hub and spoke route network structure. Second, the machine learning techniques demonstrated in this paper are only analyzed with a data set of weather-delayed flight schedules that represent the Weather functional domain. Hence, in a sequel to this paper, we extensively discuss the processes and routines used to obtain high fidelity AI models through analysis of the performance of machine learning techniques on multiple data sets that represent different functional domains in the AOCC.  

\section*{Acknowledgement}
The authors would like to thank Blair Reeves, Chien Yu Chen, Kevin Wiecek, Jeff Agold, Dave Harrington, Rick Dalton, and Phil Beck, at Southwest Airlines Network Operations Control (SWA-NOC), for their expert inputs in abstracting the data used for this work.

\section*{Conflict of Interest}
All authors have no conflict of interest to report.

\bibliography{mendeley}

\end{document}